# Rare-earth-mediated opto-mechanical system in the reversed dissipation regime

Ryuichi Ohta[1], Loïc Herpin[1], Victor M. Bastidas[1, 2],
Takehiko Tawara[1, 3], Hiroshi Yamaguchi[1], and Hajime Okamoto[1]

[1] *NTT Basic Research Laboratories, NTT Corporation*
[2] *NTT Research Center for Theoretical Quantum Physics, NTT Corporation*
[3] *NTT Nanophotonics Center, NTT Corporation*
*3-1 Morinosato Wakamiya, Atsugi-shi, Kanagawa 243-0198, Japan*

Strain-mediated interaction between phonons and telecom photons is demonstrated using excited states of erbium ions embedded in a mechanical resonator. Owing to the extremely long-lived nature of rare-earth ions, the dissipation rate of the optical resonance falls below that of the mechanical one. Thus, a "reversed dissipation regime" is achieved in the optical frequency region. We experimentally demonstrate an opto-mechanical coupling rate $g_0 = 2\pi \times 21.7$ Hz, and numerically reveal that the interaction causes stimulated excitation of erbium ions. Numerical analyses further indicate the possibility of $g_0$ exceeding the dissipation rates of erbium and mechanical systems, thereby leading to single-photon strong coupling. This strain-mediated interaction moreover involves the spin degree of freedom, and has a potential to be extended to highly-coherent opto-electro-mechanical hybrid systems in the reversed dissipation regime.

Interactions between electromagnetic and acoustic waves have been investigated through various optical resonances, such as optical cavities [1-4], microwave circuits [5-8], and solid-state two-level systems [9-16], incorporated in mechanical resonators via radiation pressure, capacitive force, and strain effects. Recently, these opto-mechanical interactions have attracted much attention in diverse fields ranging from quantum information to nonlinear optics. Sideband cooling of the mechanical mode to its ground state [1], coherent energy transfer between photons and phonons [2, 3, 6], and generation of their entangled states [4] have been demonstrated in the past decade.

In these systems, the dissipation rates of the optical and mechanical resonances are of crucial importance to determining the dynamics of the photons and phonons. For instance, to optically cool (readout) the mechanical motion (phonon state), the dissipation rate of the optical resonance should be larger than that of the mechanical one ($\gamma_{opt} > \gamma_m$). This regime is the conventional situation in cavity opto-mechanical systems. In contrast, the opposite situation ($\gamma_{opt} < \gamma_m$) inverts the roles of photons and phonons. In this regime, the optical resonance and its intra-photon are dominantly affected by the dynamic back-action of the opto-mechanical interaction, which leads to quantum-limited amplification and self-oscillation of photons [8, 17], entangled-photon generation [18], and reservoir engineering of photons for nonreciprocal manipulations [19, 20]. So far, this reversed dissipation regime has been realized in the microwave region [8]. However, it is difficult to achieve in the optical region, including telecom wavelengths which enable long-distance communication. The main obstacle is the large energy difference between photons and phonons. Although high-$Q$ optical cavities have been developed, their dissipation rates are of the order of a megahertz, i.e., much higher than mechanical dissipation rates, which typically range from a hertz to a kilohertz. Extremely long-lived optical resonances are required to interact with the mechanical degree of freedom.

In this paper, we demonstrate strain-mediated opto-mechanical interaction between erbium (Er) ions and a $Y_2SiO_5$ (YSO) mechanical resonator. The excited states of the Er ions play the role of the optical resonance whose dissipation rate $\gamma_{Er}$ falls below $2\pi \times$ 99 Hz (corresponding $Q > 10^{13}$) owing to the 4f-4f transition of rare-earth ions. This dissipation rate is five orders of magnitude lower than those of optical cavities [1-4] and other two-level systems [10-16], and one order of magnitude lower than that of the mechanical resonator in this study, thus allowing for the appearance of the reversed dissipation regime at telecom wavelengths ( $\gamma_{Er} < \gamma_m \ll \omega_m \ll \omega_{Er}$ ). The coupling rate $g_0 = 2\pi \times 21.7$ Hz was determined from stroboscopic photoluminescence excitation (PLE) measurements, which probed the transition energies of Er ions modulated by mechanical deformation. We theoretically investigated this opto-mechanical system by using the master equation approach and found that the strain-mediated interaction caused stimulated excitation of Er ions with a blue-detuned pump. Further numerical simulations indicated that $g_0$ could be enhanced to exceed the dissipation rates of the optical and mechanical resonances, leading to strong coupling between a single photon and phonon ($\gamma_{Er} < \gamma_m < g_0$). These results pave the way to nonlinear manipulation and reservoir engineering of photons, as well as highly coherent hybrid systems in the reversed dissipation regime.

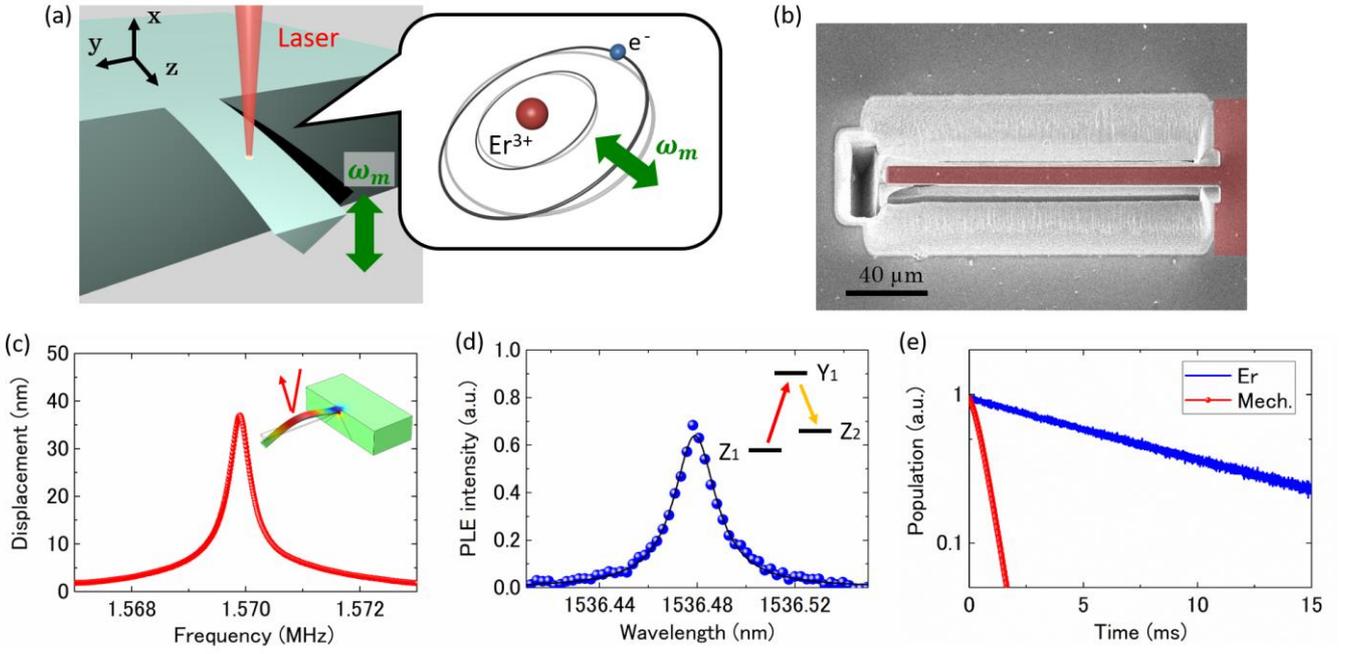

FIG. 1. (a) Schematic image of the strain-mediated opto-mechanical interaction of Er ions in a mechanical resonator. The mechanical vibration modulates the transition energy of the Er ions. (b) SEM image of Er:YSO mechanical resonator (red shaded region). (c) Measured frequency response of the second-order flexual mode of the resonator. The inset shows the shape and strain distribution of this mode calculated by FEM. (d) PLE spectrum of the $Y_1$-$Z_1$ transition of the Er ions embedded in the resonator. The inset shows the energy diagram of this transition. (e) Energy decay of the excited state of Er ions and the mechanical mode. The dissipation rate of the Er ions is one order of magnitude lowere than that of the mechanical mode.

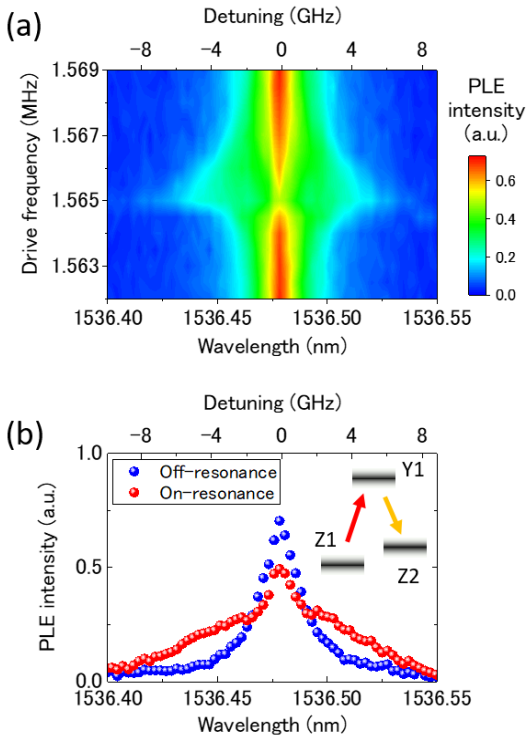

FIG. 2. (a) Drive frequency dependence of PLE spectra taken with continuous excitation of the pump laser. (b) PLE spectra at on- (red) and off-resonance (blue). The inset is a schematic image of the energy diagram with mechanical vibration.

The Er ions in this study were uniformly doped in bulk YSO crystals (provided by Scientific Materials). The concentration of Er ions was 0.1 %. The degeneracy of each energy level is lifted because of the crystal fields of YSO. Mechanical deformation of YSO geometrically perturbs its crystal field and the transition energies between $I_{15/2}$ and $I_{13/2}$ levels, which is the origin of the strain-mediated opto-mechanical interaction in this scheme [Fig. 1(a)]. Static strain effects of Er:YSO have been studied with co-doping of other rare-earth materials, such as Eu [21] and Sc [22]. These auxiliary impurities randomly generate local strain and increase the inhomogeneous linewidths of the Er ions. Previous experiments have demonstrated that the applied strain does not degrade the coherence of the excited states of the Er ions [21]. Therefore, small dephasing rates ($\gamma_2 < 1$ kHz [21, 23]) can be obtained under strain.

We fabricated mechanical resonators by using angled focused ion beam (FIB) milling [24, 25]. Figure 1(b) is a scanning electron microscope (SEM) image of a fabricated resonator taken from the top. Length, width, and milling angle were 160 μm, 16 μm, and 45 degrees, respectively. The length, width, and height were directed to the $D_1$-, $D_2$-, $b$-axes of the crystal. The sample was mounted on a piezo actuator, which electrically drove the mechanical motion, and it was cooled to a temperature of 4 K.

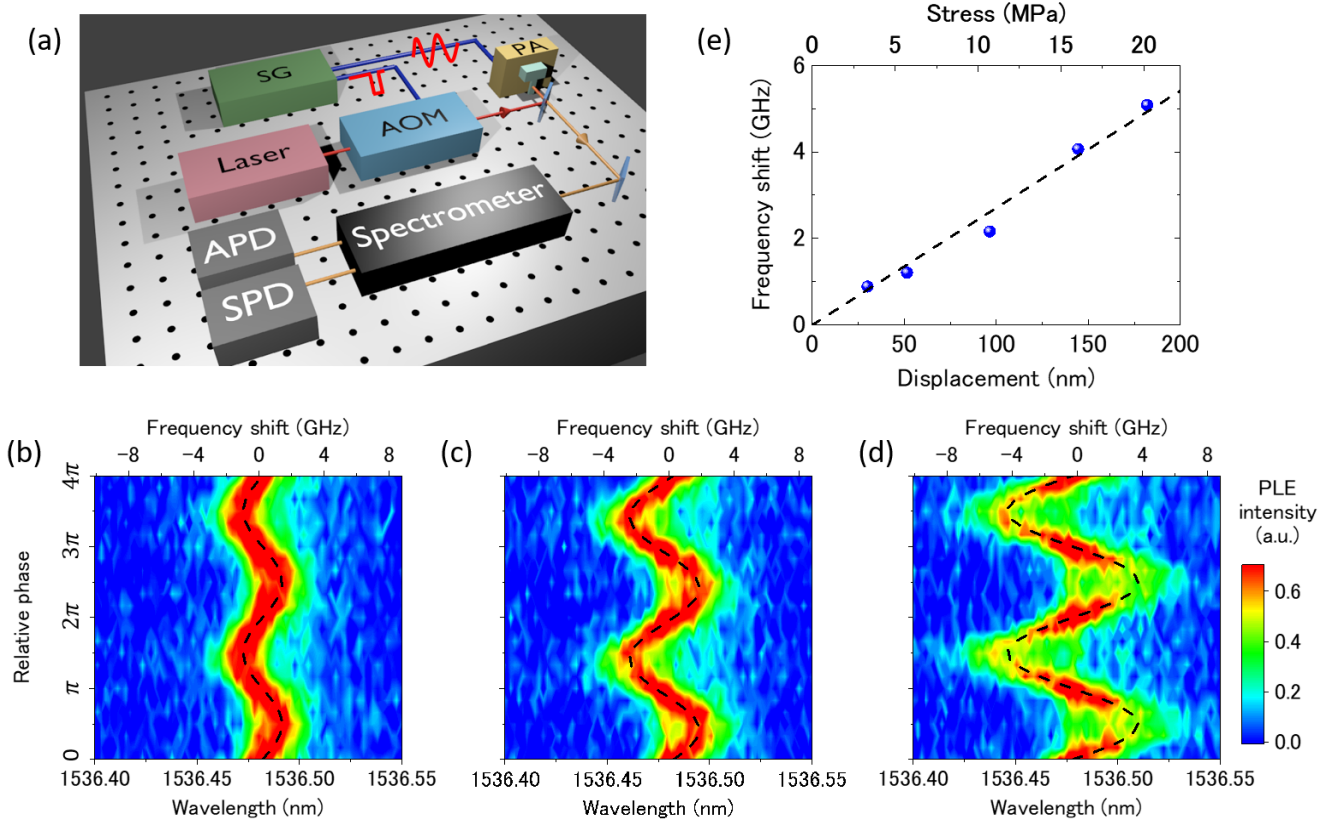

FIG. 3. (a) Schematic image of storoboscopic measurement. AOM: Acousto optical modulator, SG: Signal generator, PA: Piezo actuator, APD: Avalanche photodiode, SPD: Single photon detector. (b-d) Stroboscopic PLE spectra at drive voltages of 3 (b), 5 (c) and 10 $V_{pp}$ (d), where the measured displacements ($x_{pp}$) were 103, 193, 289 nm, respectively. The vertical axes correspond to the relative phase of the pump laser and mechanical motion. Dashed lines are curves fitted with Eq. 1. (e) Mechanical displacement and stress dependences of the frequency shifts of the transition energy of Er ions fitted by a linear function (dashed line).

The mechanical properties of the fabricated resonators were measured with a HeNe laser and a Doppler interferometer. In the following measurements, we electrically drove the second-order flexural mode of the resonator [11], whose maximum strain is located at the mid-point. Figure 1(c) shows the frequency response of this mode measured at the mid-point of the resonator. The mechanical quality factor was 2,500.

The optical properties of the excited states of the Er ions were investigated through PLE measurements. We excited the $Y_1$-$Z_1$ transition (1536.4 nm) of the doped Er ions at site 1 of the crystal and measured the luminescence from their $Y_1$-$Z_2$ transition (1546.5 nm). The laser was aligned along the $b$-axis and was focused on the mid-point and the top surface of the resonator. Figure 1(d) depicts the PLE spectrum of the Er ions embedded in the resonator without mechanical vibration. The linewidth was 2.66 GHz, which indicated intrinsic inhomogeneous broadening of the transition energies.

The dissipation rates of the Er ions and the mechanical resonance were independently evaluated with ring-down measurements. Figure 1(e) shows the energy decays of the excited state of the Er ions and the second-order flexural mode of the resonator. The decay times of the Er ions and the mechanical mode were 10.1 ms and 0.56 ms, respectively, where the corresponding dissipation rates were $\gamma_{Er} = 2\pi \times 99$ Hz and $\gamma_m = 2\pi \times 1.79$ kHz. The dissipation rate of Er ions was one order of magnitude lower than that of the mechanical mode ($\gamma_{Er} < \gamma_m$).

We experimentally demonstrated the strain-mediated opto-mechanical interaction in this reversed dissipation regime by making PLE measurements under mechanical vibration. First, we continuously measured the PLE spectra at a drive voltage of 20 $V_{pp}$, which generated a peak-to-peak mechanical displacement $x_{pp} = 357$ nm at the resonance frequency. The drive frequency dependences shown in Fig. 2 reveal that the vibrational strain significantly broadened the PLE spectra. This peak broadening was due to the time-averaging of the continuously modulated transition energy of Er ions. The full-width at half-maximum (FWHM) of this broadening was 10.5 GHz.

To evaluate the opto-mechanical coupling factors, we performed stroboscopic PLE measurements. In this scheme, the intensity of the excitation laser was modulated to define the pulse shape, whose repetition was synchronized to the mechanical resonance frequency [Fig. 3(a)]. Therefore, by changing the relative phase of the pump pulse and mechanical motion, we could measure the PLE spectra at arbitrary times of the mechanical motion. Figures 3(b)-3(d) show the stroboscopic PLE spectra at different drive voltages, $V_d$ =3, 5, and 10 $V_{pp}$, where the measured $x_{pp}$ were respectively 103, 193, and 289 nm. The mechanical strain clearly modulated the transition energies of the Er ions. The sinusoidally modulated spectra, instead of the peak broadening, indicate that the excitation laser was well focused on the top surface of the resonator. We evaluated the dispersive opto-mechanical coupling factors from the peak fittings with Eq. 1 in Figs. 3(b-d).

$$\omega_{Er} = \omega_0 + \frac{G_{disp}x_{pp}}{2}\sin(\omega_m t + \theta_0) \quad (1)$$

Here, $\omega_0$ is the steady-state transition energy of the Er ions, $G_{disp}$ is the coupling factor normalized by the mechanical displacement, $\omega_m$ represents the resonance frequency of the mechanical mode, and $\theta_0$ is the phase difference between the pump laser and the mechanical motion. Figure 3(e) shows the displacement and corresponding stress dependences of the frequency shifts, where the corresponding stress was numerically calculated with the finite element method (FEM). The linear dependence of the energy shift gives $G_{disp}$ = 27 MHz/nm. To compare the magnitude of the opto-mechanical interaction of Er ions to those of other two-level systems, we also derived the structural independent coupling factor $G_{stress}$, which corresponds to the energy shift normalized by stress. For Er ions, $G_{stress}$ is 243 Hz/Pa, which is as large as those of NV centers (465 Hz/Pa [26]), while the dissipation rate of the excited states of Er ions is six orders of magnitude smaller than that of NV centers. The opto-mechanical coupling rate $g_0$ in this system is derived from the energy shift caused by a single phonon fluctuation.

$$g_0 = G_{disp}x_{zpf} \quad (2a)$$

$$x_{zpf} = \sqrt{\frac{\hbar}{2m_{eff}\omega_m}} \quad (2b)$$

Here, $x_{zpf}$ and $m_{eff}$ are the zero-point fluctuation and effective mass of the resonator. Thus, we experimentally obtained $g_0 = 2\pi \times 21.7$ Hz for our opto-mechanical system.

As discussed in Refs. 8 and 17, the dispersive opto-mechanical interaction in the reversed dissipation regime modifies the susceptibility of the optical resonance and leads to amplification and lasing of the cavity photons with a blue-detuned pump. Here, we theoretically discuss the back-action effects of the dispersive interaction between an ensemble Er ions and a mechanical resonator by using the master equation [27-29]. In this model, we assume that the mechanical motion collectively interacts with the ensemble Er ions, that the occupancy of their excited states is much lower than unity, and that the inhomogeneity of $\omega_{Er}$ becomes small enough to use a blue-detuned pump [27].

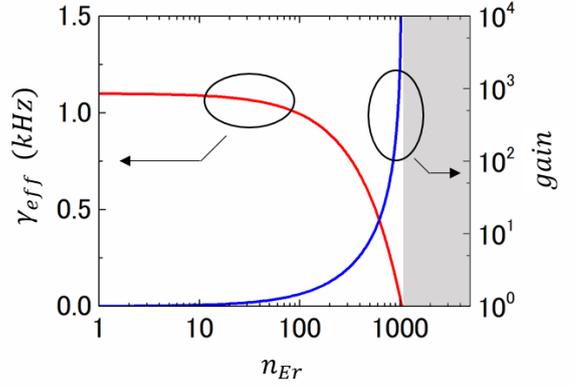

FIG. 4. Gain and effective damping rate as a function of $n_{Er}$ with blue-detuned pump. Shaded area indicates the region of the self-sustained oscillation.

$$\widehat{H}_{OM} = -\Delta_0 \hat{a}^\dagger \hat{a} + \omega_m \hat{b}^\dagger \hat{b} - g_0 \hat{a}^\dagger \hat{a}(\hat{b}^\dagger + \hat{b}) + \Omega\sqrt{N}(\hat{a}^\dagger + \hat{a}) \quad (3a)$$

$$\dot{\hat{\rho}} = i[\widehat{H}_{OM}, \hat{\rho}] + \frac{\gamma_1}{2}\mathcal{D}(\hat{a})\hat{\rho} + \frac{\gamma_2}{2}\mathcal{D}(\hat{a}^\dagger \hat{a})\hat{\rho} + \frac{\gamma_m}{2}\mathcal{D}(\hat{b})\hat{\rho} \quad (3b)$$

Here, $\hat{a}$ ($\hat{a}^\dagger$) and $\hat{b}$ ($\hat{b}^\dagger$) are the annihilation (creation) operators of Er ions and phonons, $\Delta_0 = \omega_L - \omega_{Er}$ is the detuning between the pump and transition frequency, $N$ is the number of interacted Er ions, $\Omega$ is the pump amplitude, $\hat{\rho}$ is the density matrix, $\gamma_1 = 2\pi \times 99$ Hz and $\gamma_2 = 2\pi \times 1$ kHz [21, 23] are the dissipation and dephasing rates of Er ions, and $\mathcal{D}(\hat{O})\hat{\rho} = 2\hat{O}\hat{\rho}\hat{O}^\dagger - \hat{O}^\dagger\hat{O}\hat{\rho} - \hat{\rho}\hat{O}^\dagger\hat{O}$ for a given operator $\hat{O}$. Under a blue-detuned pump, the effective damping rate of the Er ions ($\gamma_{eff}$) decreases as a function of the number of excited ions ($n_{Er} = |\bar{a}|^2$) [Fig. 4]. This opto-mechanical interaction simultaneously amplifies the excitation efficiency of Er ions (gain), which leads to self-sustained oscillation of Er ions in analogy with the opto-mechanically induced instabilities [8, 17].

The interaction between the low-loss optical and mechanical resonances has prospects for quantum and many-body physics. As described in Eq. 2, $g_0$ in this system increases as the effective mass of the resonator decreases. We numerically derived that $g_0$ reaches $2\pi \times 2$ kHz and exceeds both dissipation rates for a resonator whose length and width are 20 μm and 1 μm [27], which are still large enough to be fabricated with this method. In this regime, a single photon can be strongly coupled to a single phonon, whereas the reported opto-mechanical strong couplings have been achieved with multiple photons to amplify the effective coupling rate [5, 30, 31]. This single-photon strong coupling enables one to construct multipartite entangled spin systems in a solid-state platform [32].

In conclusion, we experimentally demonstrated a strain-mediated interaction between the optical resonance of Er ions and a mechanical resonator. The long-lived excited states of Er ions enabled us to reach the reversed dissipation regime at telecom wavelengths, which applies back-action on the optical resonance and amplifies the excitation rate of the Er ions. The extremely small dissipation of Er ions has a potential to provide strong coupling between a single photon and phonon. Our results offer new directions for research into opto-mechanical physics and will pave the way to coherent manipulation of photons, phonons, and even electrons in the reversed dissipation regime.

We thank M. Hiraishi, T. Inaba, X. Xu, M. Asano, H. Toida, M. Ono, S. Kita for technical supports and fruitful discussions.

# Supplemental Material for "Rare-earth-mediated opto-mechanical system in the reversed dissipation regime"

Ryuichi Ohta[1], Loïc Herpin[1], Victor M. Bastidas[1,2],
Takehiko Tawara[1,3], Hiroshi Yamaguchi[1], and Hajime Okamoto[1]

[1] NTT Basic Research Laboratories, NTT Corporation,
[2] NTT Research Center for Theoretical Quantum Physics, NTT Corporation
[3] NTT Nanophotonics Center, NTT Corporation
3-1 Morinosato Wakamiya, Atsugi-shi, Kanagawa 243-0198, Japan

## MASTER EQUATION APPROACH TO DESCRIBING THE RARE-EARTH-MEDIATED OPTO-MECHANICAL SYSTEM

In this section, we describe the system Hamiltonian of Er ions embedded in a mechanical resonator by using the master equation approach [28]. We suppose that an ensemble of $N$ Er ions is externally pumped and is collectively coupled to a mechanical mode. Thus, the system is described by the Hamiltonian,

$$\hat{H}_{OM} = -\sum_{j=1}^{N}(\omega_L - \omega_j^{Er})|e\rangle\langle e|_j + \omega_m \hat{b}^\dagger \hat{b}$$
$$-g_0 \sum_{j=1}^{N}|e\rangle\langle e|_j (\hat{b} + \hat{b}^\dagger) + \Omega \sum_{j=1}^{N}(|e\rangle\langle g|_j + |g\rangle\langle e|_j) \quad (1)$$

where $\omega_j^{Er}$ is the resonance frequency of the $j$-th Er ions and $\omega_L$ is the pump frequency. To simplify the discussion, we will neglect the interactions between individual Er ions, and the existence of thermal phonons giving additional noise to the system, which becomes crucially important in the quantum regime [28]. In this analysis, we assume a homogeneous ensemble such that $\omega_j^{Er} = \omega^{Er}$, so that we can define collective spin operators.

Next, we map the system to a bosonic mode in the low excitation regime $\langle \hat{a}^\dagger \hat{a}\rangle \ll N$, by using the Holstein-Primakoff representation of the angular momentum algebra $\hat{J}^z = \hat{a}^\dagger\hat{a} - N/2$, $\hat{J}^+ = \hat{a}^\dagger(\sqrt{N - \hat{a}^\dagger\hat{a}})$, and $\hat{J}^- = (\sqrt{N - \hat{a}^\dagger\hat{a}})\hat{a}$.

$$\sum_{j=1}^{N}|e\rangle\langle e|_j = \frac{1}{2}\sum_{j=1}^{N}(\sigma_j^z + 1) = \hat{J}^z + \frac{N}{2} = \hat{a}^\dagger\hat{a} \quad (2a)$$

$$\sum_{j=1}^{N}|e\rangle\langle g|_j = \sum_{j=1}^{N}\sigma_j^+ = \hat{J}^+ \approx \sqrt{N}\hat{a}^\dagger \quad (2b)$$

$$\sum_{j=1}^{N}|g\rangle\langle e|_j = \frac{1}{2}\sum_{j=1}^{N}\sigma_j^- = \hat{J}^- \approx \sqrt{N}\hat{a} \quad (2c)$$

The assumption of a low excitation regime is justified by our experimental results on the linear pump-power dependence of the PL intensity. After this transformation, the system Hamiltonian has the same form as the reported opto-mechanical Hamiltonian [17]

$$\hat{H}_{OM} = -\Delta_0 \hat{a}^\dagger\hat{a} + \omega_m \hat{b}^\dagger\hat{b} - g_0 \hat{a}^\dagger\hat{a}(\hat{b} + \hat{b}^\dagger)$$
$$+ \Omega\sqrt{N}(\hat{a} + \hat{a}^\dagger) \quad (3)$$

where $\Delta_0 = \omega_L - \omega^{Er}$ is the detuning. To account for both the dissipation and dephasing of the system, we use the framework of the master equation for the density matrix of the system.

$$\frac{d\hat{\rho}}{dt} = i[\hat{H}_{OM}, \hat{\rho}] + \frac{\gamma_1}{2}\mathcal{D}(\hat{a})\hat{\rho} + \frac{\gamma_2}{2}\mathcal{D}(\hat{a}^\dagger\hat{a})\hat{\rho} + \frac{\gamma_m}{2}\mathcal{D}(\hat{b})\hat{\rho} \quad (4)$$

$\gamma_1$ and $\gamma_2$ are the dissipation and dephasing rates of Er ions, and $\mathcal{D}(\hat{O})\hat{\rho} = 2\hat{O}\hat{\rho}\hat{O}^\dagger - \hat{O}^\dagger\hat{O}\hat{\rho} - \hat{\rho}\hat{O}^\dagger\hat{O}$ for a given operator $\hat{O}$. We neglect the dephasing of the mechanical resonator. From the master equation, we can obtain the equations of motions of $\hat{a}$ and $\hat{b}$.

$$\frac{d\hat{a}}{dt} = i\Delta_0 \hat{a} + ig_0\hat{a}(\hat{b} + \hat{b}^\dagger) - \frac{\gamma}{2}\hat{a} + \sqrt{\gamma_1}a_{in} \quad (5a)$$

$$\frac{d\hat{b}}{dt} = -i\omega_m \hat{b} + ig_0\hat{a}^\dagger\hat{a} - \frac{\gamma_m}{2}\hat{b} \quad (5b)$$

$\gamma = \gamma_1 + \gamma_2$, and $\sqrt{\gamma_1}a_{in} = -i\Omega\sqrt{N}$. Net, we follow the procedure reported in Ref. 17 to derive the susceptibility of this system. $\hat{a}$ ($\hat{b}$) is decomposed into the mean value $\bar{a}$ ($\bar{b}$) and the fluctuation $\delta\hat{a}$ ($\delta\hat{b}$). Neglecting nonlinear terms, they are described as

$$\frac{d\bar{a}}{dt} = i\Delta\bar{a} + ig_0\bar{a}(\bar{b} + \bar{b}^\star) - \frac{\gamma}{2}\bar{a} + \sqrt{\gamma_1}\bar{a}_{in} \quad (6a)$$

$$\frac{d\bar{b}}{dt} = -i\omega_m \bar{b} + ig_0|\bar{a}|^2 - \frac{\gamma_m}{2}\bar{b} \quad (6b)$$

$$\frac{d\delta\hat{a}}{dt} = i\Delta\delta\hat{a} + iG(\delta\hat{b} + \delta\hat{b}^\dagger) - \frac{\gamma}{2}\delta\hat{a} + \sqrt{\gamma_1}\delta\hat{a}_{in} \quad (6c)$$

$$\frac{d\delta\hat{b}}{dt} = -i\omega_m \delta\hat{b} + iG(\delta\hat{a} + \delta\hat{a}^\dagger) - \frac{\gamma_m}{2}\delta\hat{b} + \sqrt{\gamma_m}\delta\hat{b}_{in} \quad (6d)$$

where $\Delta = \Delta_0 + g_0(\bar{b} + \bar{b}^\star)$, $G = g_0\bar{a}$, and $|\bar{a}|^2 = \bar{a}\bar{a}^\star$.

The susceptibility is described by the transfer matrix of $\delta\hat{a}$ in the Fourier domain.

$$\begin{pmatrix} \delta\hat{a}_{(\omega)} \\ \delta\hat{a}^\dagger_{(\omega)} \end{pmatrix}$$
$$= \frac{\sqrt{\gamma_1}}{\mathcal{N}_{(\omega)}} \begin{pmatrix} \chi_{Er}^{*-1}{}_{(-\omega)} - i\Sigma_{(\omega)} & -i\Sigma_{(\omega)} \\ +i\Sigma_{(\omega)} & \chi_{Er}^{-1}{}_{(\omega)} - i\Sigma_{(\omega)} \end{pmatrix} \begin{pmatrix} \delta\hat{a}_{in(\omega)} \\ \delta\hat{a}^\dagger_{in(\omega)} \end{pmatrix} \quad (7a)$$
$$+ \frac{iG\sqrt{\gamma_m}}{\mathcal{N}_{(\omega)}} \begin{pmatrix} \chi_{Er}^{*-1}{}_{(-\omega)}\chi_{m(\omega)} & \chi_{Er}^{*-1}{}_{(-\omega)}\chi_{m(-\omega)}^* \\ -\chi_{Er}^{-1}{}_{(\omega)}\chi_{m(\omega)} & -\chi_{Er}^{-1}{}_{(\omega)}\chi_{m(-\omega)}^* \end{pmatrix} \begin{pmatrix} \delta\hat{b}_{in(\omega)} \\ \delta\hat{b}^\dagger_{in(\omega)} \end{pmatrix}$$

$$\mathcal{N}_{(\omega)} = \chi_{Er}^{-1}{}_{(\omega)}\chi_{Er}^{*-1}{}_{(-\omega)} - 2\Delta\Sigma_{(\omega)} \quad (7b)$$

$$\Sigma_{(\omega)} = -iG^2\left(\chi_{m(\omega)} - \chi_{m(-\omega)}^*\right) \quad (7c)$$

$$\chi_{Er}^{-1}{}_{(\omega)} = \frac{\gamma}{2} - i(\omega + \Delta) \quad (7d)$$

$$\chi_m^{-1}{}_{(\omega)} = \frac{\gamma_m}{2} - i(\omega - \omega_m) \quad (7e)$$

The (1, 1) component of the first term indicates the susceptibility of the optical resonance with this strain-mediated opto-mechanical interaction. We define the effective damping rate ($\gamma_{eff}$), which include mechanical self-energy ($\Sigma_{(\omega)}$) [S1]. We also define the *gain* as the ratio of the transition coefficients with and without the pump.

$$\gamma_{eff(\omega)} = \gamma + 2\text{Im}[\Sigma_{(\omega)}] \quad (8a)$$

$$Gain_{(\omega)} = \frac{1}{\mathcal{N}_{(\omega)}\chi_{Er(\omega)}}\left(\chi_{Er}^{*-1}{}_{(-\omega)} - i\Sigma_{(\omega)}\right) \quad (8b)$$

Under the resonantly blue-detuned condition, $\Delta = \omega_m$, $\gamma_{eff}$ and *Gain* can be described as follows, where we assume $\gamma \ll \omega_m$.

$$\gamma_{eff} = \gamma - \frac{G^2}{\gamma_m} \quad (9a)$$

$$Gain = \frac{1}{1 - \frac{G^2}{\gamma_m\gamma}} \quad (9b)$$

## EXPERIMENTAL FEASIBILITY OF THE BLUE-DETUNED PUMP TO OVERCOME INHOMOGENEOUS BROADENING

As discussed in the main text, the blue-detuned pump results in amplification of the phonon-mediated excitation of Er ions. However, owing to the large inhomogeneity of $\omega_{Er}$, it is hard to selectively excite the blue-detuned sideband in the current device. In this section, we discuss how to achieve the resolved-sideband regime in this system, which requires that $\omega_m$ is not only higher than $\gamma_{Er}$ but also higher than the inhomogeneous broadening of $\omega_{Er}$ ($\Gamma_{Er}$). One way is to reduce the inhomogeneity is to decrease the density of Er ions. $\Gamma_{Er}$ becomes lower than 200 MHz at a density of 0.001 %, and it falls below 15 MHz if the Er ions are doped in symmetric host materials such as LiYF$_4$ [S2]. In this case, the resolved-sideband regime can be achieved with conventional mechanical resonators whose $\omega_m$ is several tens of MHz. Alternatively, high-frequency acoustic modes enable the resolved-sideband regime even with the current $\Gamma_{Er}$. Surface and bulk acoustic waves generate phonon-sidebands of the transition energies of quantum dots [13] and NV centers [14], whose homogeneous linewidths are a few GHz. Another approach is to apply the blue-detuned pump by means of spectral hole-burning. This enables one to selectively control the population of excited states to a spectral resolution less than 1 MHz. Therefore, an appropriate pump-pulse sequence will simultaneously create excited ions and phonons [S3].

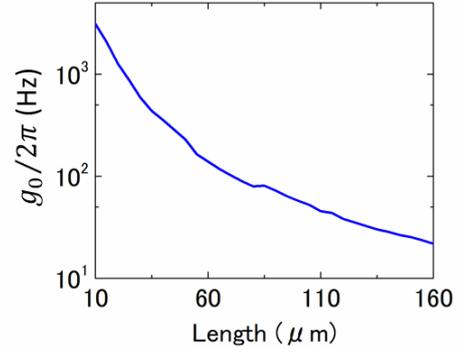

FIG. S1. Structural dependence of the coupling rate ($g_0$) calculated with FEM method. The length and width change at the same ratio with fixed $G_{stress}$.

## DESIGN OF THE RESONATOR IN THE STRONG COUPLING REGIME

We numerically investigated how large $g_0$ can be practically available in this scheme. To evaluate the general tendency, we varied the length and width of the resonator at the same ratio, while $G_{stress}$ was fixed to 243 Hz/Pa. We calculated $\omega_m$ and $m_{eff}$ of the 2$^{\text{nd}}$ flexural mode, and the ratio of the stress to $x_{pp}$ at the mid-point of the resonator. As shown in Fig. S1, $g_0$ abruptly increases as the length decreases, and reaches $2\pi \times 3$ kHz for a 10-µm-long and 1-µm-wide resonator.

Next, we optimized the design to reach the single-photon strong coupling regime, where $g_0$ exceeds $\gamma_{Er}/4$ and $\gamma_m/4$. To investigate the structural dependence of $\gamma_m$, we fabricated a smaller resonator whose length and width were 60 and 10 µm. The measured $Q$-factor was almost the same as that of the resonator described in the main text. Therefore, we assumed that the mechanical $Q$-factor does not significantly change as $m_{eff}$ decreases, and thus $\gamma_m$ is proportional to $\omega_m$, while $\gamma_{Er}$ does not change. From the numerical calculation, a 20-µm-long and 1-µm-wide mechanical resonator gives $g_0 = 2\pi \times 2$ kHz and $\omega_m = 2\pi \times 6$ MHz corresponding to $\gamma_m = 2\pi \times 6.8$ kHz. Thus, a single-photon strong coupling regime ($g_0 > \gamma_{Er}/4$, $\gamma_m/4$) could be achieved with a strain-mediated opto-mechanical interaction via Er ions.